\documentstyle[12pt]{article}
\begin{document}

\tolerance=5000
\def\pp{{\, \mid \hskip -1.5mm =}}
\def\cL{{\cal L}}
\def\be{\begin{equation}}
\def\ee{\end{equation}}
\def\bea{\begin{eqnarray}}
\def\eea{\end{eqnarray}}
\def\tr{{\rm tr}\, }
\def\nn{\nonumber \\}
\def\e{{\rm e}}
\def\D{{D \hskip -3mm /\,}}

\def\SEH{S_{\rm EH}}
\def\SGH{S_{\rm GH}}
\def\AdS5{{{\rm AdS}_5}}
\def\S4{{{\rm S}_4}}
\def\gfv{{g_{(5)}}}
\def\gfr{{g_{(4)}}}
\def\SC{{S_{\rm C}}}
\def\RH{{R_{\rm H}}}

\def\wlBox{\mbox{
\raisebox{0.1cm}{$\widetilde{\mbox{\raisebox{-0.1cm}\fbox{\ }}}$}}}
\def\htBox{\mbox{
\raisebox{0.1cm}{$\hat{\mbox{\raisebox{-0.1cm}{$\Box$}}}$}}}
\def\K{\left(k - (d-2) \e^{-2\lambda}\right)}

\  \hfill 
\begin{minipage}{3.5cm}
%OCHA-PP-189 \\
%NDA-FP-63 \\
%your preprint No. \\
April 2002 \\
%hep-th/yymmxxx \\
\end{minipage}

\vfill

\begin{center}
{\large\bf  (Anti-) de Sitter Black Holes in higher derivative gravity and
dual Conformal Field Theories}

\vfill

{\sc Shin'ichi NOJIRI}\footnote{nojiri@cc.nda.ac.jp} 
and {\sc Sergei D. ODINTSOV}$^{\spadesuit}$\footnote{
odintsov@mail.tomsknet.ru}

\vfill

{\sl Department of Applied Physics \\
National Defence Academy, 
Hashirimizu Yokosuka 239, JAPAN}

\vfill

{\sl $\spadesuit$ 
Tomsk Pedagogical University, 634041 Tomsk, RUSSIA}

\vfill

{\bf ABSTRACT}

\end{center}
Thermodynamics of five-dimensional Schwarzschild Anti-de Sitter (SAdS) and
SdS black holes in
the particular model of higher derivative gravity is considered.
The free energy, mass (thermodynamical energy) and entropy are evaluated.
There exists the parameters region where BH entropy is zero or negative.
 The arguments are given that corresponding BH solutions are not stable.
We also present the FRW-equations of motion of time (space)-like branes 
in SAdS or SdS  BH  background. The properties of dual CFT are discussed
and it is shown that it has zero Casimir energy 
when BH entropy (effective gravitational constant) is zero. The
Cardy-Verlinde formula for CFT dual to SAdS or SdS BH is found in the
universal form.

\newpage

\section{Introduction}

AdS/CFT correspondence \cite{AdS} and proposed dS/CFT correspondence 
\cite{strominger,hull,ds2}(for recent study see refs.\cite{ds3a,ogushi,others,
cvetic} and references therein) provided quite deep motivation to 
study AdS and dS black hole thermodynamical properties, 
respectively. The additional related motivation comes from the 
brane-world approach.
It is quite well-known that five-dimensional AdS Einstein gravity 
gives the description of dual four-dimensional CFT in the leading 
order of large $N$ approximation.
Higher derivative gravity is natural framework to investigate 
the non-perturbative properties of dual CFT in next-to-leading order.
The study of AdS and dS black holes in five-dimensional classical 
higher derivative gravity \cite{NOOr2a} demonstrated the possibility 
of negative entropy black holes \cite{cvetic}. This is quite unusual 
situation which is not typical in Einstein gravity. It is caused 
mainly by consideration of the higher derivative terms 
coefficients on the equal footing with gravitational constant.

 From the thermodynamical point of view the objects with negative 
entropy are not physical ones and they should not occur. Hence, 
it is expected that there should be some mechanism to suppress 
the appearence of such objects.
Currently, one can provide the following points of view to 
resolve this problem.

\noindent
1. Superstring/M-theory compactifications should predict the physical
values of the higher derivative terms coefficients in the 
gravitational action. Presumably, the corresponding values are 
such that region of parameters admitting negative entropy black 
holes is never realized  in physical world. This point of view was 
suggested in ref.\cite{cvetic}. 
However, so far there is no explicit realization of  realistic
compactified theory.

\noindent
2. Quantum higher derivative gravity which is renormalizable in four
dimensions (for review, see \cite{BOS}) is famous due to non-unitarity
problem. This may indicate that higher derivative terms should be
accounted only at microscopic QG scale. In other words, strange 
classical effects of higher derivative gravity should be 
suppressed quantum mechanically. 

\noindent
3. The negative entropy black holes may indicate to the breaking 
of some physical picture. For example, the appearence of such 
objects may indicate to the possibility of some new sort of 
phase transition (say, between de Sitter and Anti-de Sitter 
black holes \cite{cvetic}). If so, some manifestation of such
phase transition should be seen in dual CFT description.
Moreover, negative entropy black hole solutions should not be stable.
Finally, the resolution may lie in the combination of above points.

The purpose of this work is to study the thermodynamics
(entropy, free energy, mass) of five-dimensional AdS and dS black holes in
sufficiently simple model of higher derivative gravity. Nevertheless, such
model admits 
the realization of negative entropy black holes. We explicitly 
show that negative entropy black holes are not stable.
The FRW equations of motion of time (space)-like branes in SAdS or SdS 
 black holes are considered. The study of dual CFT shows that when BH
entropy is zero, the brane CFT energy is also zero. The Cardy-Verlinde
formula for (A)dS BHs is presented.. 

\section{(Anti)-de Sitter Black Hole thermodynamics}

In attempt to understand the thermodynamics of (A)dS BHs let us first
present the short review of such objects in Einstein gravity.
The 5-dimensional Einstein gravity  action is given by, 
\be
\label{E1}
S=\int d^5 x \sqrt{-g}\left\{{1 \over \kappa^2} R - \Lambda 
\right\}\ .
\ee
When $\Lambda$ is negative, the Schwarzschild anti-de Sitter 
(SAdS) spacetime is a solution of the Einstein equation:
\be
\label{E2}
ds^2_{\rm Einstein}=-\e^{2\rho}dt^2 
+ \e^{-2\rho}dr^2 + r^2 g_{ij}dx^i dx^j\ ,\quad 
\e^{2\rho}= {k \over 2} - {\mu_{\rm AdS} \over r^2} 
+ {r^2 \over l_{\rm AdS}^2}\ .
\ee
Here $g_{ij}$ is the metric of the 3-dimensional 
Einstein manifold, where the 
Ricci curvature $R_{ij}$ satisfies the relation 
$R_{ij}=kg_{ij}$. The case of $k=2$ corresponds to the unit sphere, 
$k=-2$ to the unit hyperboloid and, as a special case, $k=0$ to the 
flat space. The length parameter $l_{\rm AdS}$ is  
\be
\label{E3}
l_{\rm AdS}^2=-{\kappa^2 \Lambda \over 12}\ .
\ee
On the other hand, when $\Lambda$ is positive, the Schwarzschild- 
de Sitter (SdS) spacetime is a solution:
\be
\label{E4}
ds^2_{\rm Einstein}=-\e^{2\rho}dt^2 
+ \e^{-2\rho}dr^2 + r^2d\Omega_3^2 \ ,\quad 
\e^{2\rho}= 1 - {\mu_{\rm dS} \over r^2} - {r^2 \over l_{\rm dS}^2}\ .
\ee
Here $d\Omega_3^2$ is the metric of the unit 3-dimensional 
sphere. The length parameter is given by 
\be
\label{E5}
l_{\rm dS}^2=-l_{\rm AdS}^2={\kappa^2 \Lambda \over 12}\ .
\ee
There is a black hole horizon at
\be
\label{E6}
r^2=r^2_{\rm AdS-bh}\equiv {-{k \over 2}l_{\rm AdS}^2 
+ \sqrt{l_{\rm AdS}^4\left({k \over 2}\right)^2 
+ 4\mu_{\rm AdS} l_{\rm AdS}^2} \over 2}
\ee
for the SAdS spacetime and 
\be
\label{E7}
r^2=r^2_{\rm dS-bh}\equiv {l_{\rm dS}^2 
 - \sqrt{l_{\rm dS}^4 - 4\mu_{\rm dS} l_{\rm dS}^2} \over 2}
\ee
for the SdS spacetime. For the 
Schwarzschild-dS spacetime there is also cosmological horizon at
\be
\label{E8}
r^2=r^2_{\rm csm}\equiv {l_{\rm dS}^2 
+ \sqrt{l_{\rm dS}^4 - 4\mu_{\rm dS} l_{\rm dS}^2} \over 2}\ .
\ee
Then the corresponding Hawking temperatures are given by
\bea
\label{E9}
T_{\rm AdS-bh} = {k \over 4\pi r_{\rm AdS-bh}} 
+{r_{\rm AdS-BH} \over \pi l_{\rm AdS}^2} 
\eea
for the SAdS spacetime and 
\be
\label{E10}
T_{\rm dS-bh,csm}=\left|{\mu_{\rm dS} \over 2\pi 
r_{\rm dS-bh,csm}^3} 
 - {r_{\rm dS-bh,csm} \over 2\pi l_{\rm dS}^2}\right|
= {\sqrt{l_{\rm dS}^4 - 4\mu_{\rm dS} l_{\rm dS}^2} 
\over 2\pi l_{\rm dS}^2 r_{\rm dS-bh,csm}}
\ee
for the SdS spacetime. 
In \cite{BK,BBM}, the mass $M$ of the black hole has been 
evaluated by using the surface counterterms method. One can regard 
the mass as the thermodynamical energy. Then 
\be
\label{E11}
E_{\rm AdS}={3l_{\rm AdS}^2 V_3 \over 16\kappa^2} \left(k^2  
+ {16\mu_{\rm AdS} \over l_{\rm AdS}^2}\right)
\ee
for the SAdS spacetime. 
Here $V_3$ is a volume of 3d surface with unit radius. 
When the surface is 3d sphere one has $k=2$ and $V_3=2\pi^2$. 
On the other hand, for the SdS spacetime, one obtains 
\be
\label{E12}
E_{\rm dS}={3\pi^2 l_{\rm dS}^2 \over 2\kappa^2}
 - {6\pi^2 \mu_{\rm dS} \over \kappa^2}\ ,
\ee
which corresponds to $k=2$ and $V_3=2\pi^2$ case.
The parts independent of $\mu_{\rm AdS,dS}$ in (\ref{E11}) 
and (\ref{E12}) could 
be regarded as the Casimir energy of the bulk spacetime. 
Using the thermodynamical relation $d{\cal S}={dE \over T}$, 
we find the entropy 
\be
\label{E13}
{\cal S}_{\rm AdS}=\int {dE \over T_{\rm AdS-bh}} = {4V_{3}\pi 
r_{\rm AdS-bh}^3 
\over \kappa^2} + {\cal S}_0^{\rm AdS}\ .
\ee
for the SAdS spacetime. 
Here $S_0^{\rm AdS}$ is a constant of the integration. 
If one requires that the entropy should vanish for the 
pure AdS with $\mu_{\rm AdS}=r_{\rm AdS-bh}=0$, one arrives at 
$S_0^{\rm AdS}=0$.
It is funny that if the gravitational constant would be negative,
then the entropy also would be negative.

On the other hand, when the spacetime is asymptotically dS, 
we have 
\be
\label{E14}
{\cal S}_{\rm dS}= {4V_{3}\pi r_{\rm csm}^3 \over \kappa^2}
+ {\cal S}_0^{\rm dS} \ .
\ee
Here $S_0^{\rm dS}$ is a constant of the integration, again. 
When the spacetime is asymptotically 
de Sitter, the integration constant ${\cal S}_0^{\rm dS}$ may 
be determined by requiring that the entropy ${\cal S}_{\rm dS}$ 
should vanish in the Nariai limit $r_{\rm dS-bh,csm}^2\rightarrow 
{l_{\rm dS}^2 \over 2}$: 
\be
\label{E15}
S_0^{\rm dS}= -{2V_{3}\pi l_{\rm dS}^3 \over \sqrt{2}\kappa^2}
\ee
Now, one can construct the same BHs in higher derivative gravity.

General action of $D=d+1$ dimensional 
$R^2$-gravity with cosmological constant is:
\be
\label{vi}
S=\int d^{d+1} x \sqrt{-G}\left\{a R^2 + b R_{\mu\nu} R^{\mu\nu}
+ c  R_{\mu\nu\xi\sigma} R^{\mu\nu\xi\sigma}
+ {1 \over \kappa^2} R - \Lambda 
\right\}\ .
\ee
When $c=0$ which is the case under consideration, Schwarzschild-(anti) de
Sitter space is 
an exact solution:
\bea
\label{SAdSA}
ds^2&=&G_{\mu\nu}dx^\mu dx^\nu \nn
&=&-\e^{2\rho_0}dt^2 + \e^{-2\rho_0}dr^2 
+ r^2\sum_{i,j}^{d-1} g_{ij}dx^i dx^j\ ,\nn
\e^{2\rho}&=&{1 \over r^{d-2}}\left(-\mu_{\rm AdS} 
+ {kr^{d-2} \over d-2} + {r^d \over l_{\rm AdS}^2}\right)\ .
\eea
The curvatures have the following form:
\be
\label{cvA}
R=-{d(d+1) \over l_{\rm AdS}^2}\ ,\quad 
R_{\mu\nu}= - {d \over l_{\rm AdS}^2} G_{\mu\nu}\ .
\ee
In (\ref{SAdSA}), $\mu_{\rm AdS}$ is the parameter 
corresponding to mass and the length parameter $l_{\rm AdS}$ 
is found by solving the following equation:
\be
\label{llA}
0={d^2(d+1)(d-3) a \over l_{\rm AdS}^4} + {d^2(d-3) b \over 
l_{\rm AdS}^4} \nn
- {d(d-1) \over \kappa^2 l_{\rm AdS}^2}-\Lambda\ .
\ee
We also assume $g_{ij}$ corresponds to the Einstein manifold, 
again. When $l_{\rm AdS}^2$ is positive(negative), the spacetime is 
asymptotically anti-de Sitter (de Sitter). 
In the following, we concentrate on the case of $d=4$. 
Then Eq.(\ref{llA}) can be rewritten in the following form:
\bea
\label{llA4}
0&=&l_{\rm AdS}^4 + {12 l_{\rm AdS}^2 \over \kappa^2\Lambda} 
 - {80a + 16b \over \Lambda} \nn
&=&\left(l_{\rm AdS}^2 + {5 \over \kappa^2\Lambda}\right)^2 
 - {36 \over \kappa^4\Lambda^2} 
 - {80a + 16b \over \Lambda} \ .
\eea
Therefore if and only if
\be
\label{dSAdS1}
{36 \over \kappa^4\Lambda^2} + {80a + 16b \over \Lambda}
\geq 0\ ,
\ee
there are (real) solutions:
\be
\label{dSAdS2}
{1 \over l_{\rm AdS}^2}={{6 \over \kappa^2}\pm 
\sqrt{{36 \over \kappa^4} 
+ \Lambda (80a + 16b)} \over 80a + 16b}\ .
\ee
Eq.(\ref{llA4}) tells that the signs of solution $l_{\rm AdS}^2$ 
depend on the signs of $\kappa^2\Lambda$ and ${80a + 16b 
\over \Lambda}$. 
\begin{enumerate}
\item When $80a+16b>0$ and $\Lambda>0$ or 
$80a+16b<0$ and $\Lambda<0$, one solution for $l_{\rm AdS}^2$ 
(\ref{llA4}) is positive but another is negative. Therefore 
there are two kinds of solution, the 
 asymptotically AdS spacetime is first solution and 
another one is de Sitter space. 
\item When $80a+16b>0$ and $\Lambda<0$, both of two solutions 
for $l_{\rm AdS}^2$  (\ref{llA4}) are positive. Therefore 
these solutions correspond to anti-de Sitter space. 
\item When $80a+16b<0$ and $\Lambda>0$, both of two solutions 
for $l_{\rm AdS}^2=-l_{\rm dS}^2$  (\ref{llA4}) are negative. 
Therefore these solutions correspond to de Sitter space. 
\end{enumerate}
Since there always appear two solutions, we now investigate 
the entropy. 
The entropy can be obtained by regarding the black hole mass 
as the thermodynamical mass. The thermodynamical mass is again
obtained by using the surface counterterms method \cite{cvetic}. 
If the spacetime is asymptotically anti-de Sitter, the $(t,t)$ 
component of the surface energy momentum tensor has 
the following form for general $R^2$-gravity
\be
\label{ABC9}
T^{tt}={3 \over 4l_{\rm AdS}r^6}\left({1 \over \kappa^2} 
 - {40a \over l_{\rm AdS}^2} - {8b \over l_{\rm AdS}^2} 
 - {4c \over l_{\rm AdS}^2}\right)\ .
\ee
The black hole mass $M$ can be evaluated at the 3d 
surface where $t$ is a constant:
\be
\label{ABC10}
M=\int dx^3\sqrt{\tilde g}r^3 \e^\nu T^{tt}\left(\xi_t\right)^2
\ .
\ee
Here $\zeta^\mu$ is a unit vector parallel with the time 
coordinate $t$ and therefore $\zeta^t=\e^{-\nu}$ 
($\zeta_t=\e^\nu$) and $\zeta^\mu=\zeta_\mu=0$ ($\mu\neq t$). 
Note $\sqrt{\tilde g}\e^\nu = \sqrt{\det\,g_{mn}}$.  
Then  for the asymptotically AdS spacetime 
\be
\label{ABC11}
M_{\rm AdS}=E_{\rm AdS}
={3l_{\rm AdS}^2 V_3 \over 16} \left({1 \over \kappa^2} 
 - {40a \over l_{\rm AdS}^2}  - {8b \over l_{\rm AdS}^2} 
 - {4c \over l_{\rm AdS}^2}\right)\left(k^2  
+ {16\mu_{\rm AdS} \over l_{\rm AdS}^2}\right)
\ee
When $c=0$, in a way similar to the Einstein gravity case, we can 
obtain the expression for the entropy considering $M$ as a 
thermodynamical energy. When the spacetime is asymptotically 
anti-de Sitter, the entropy is given by
\be
\label{lll5}
{\cal S}_{\rm AdS}= {V_{3}\pi r_{\rm AdS-bh}^3 \over 2}
\left( {8 \over \kappa^2}- {320 a + 64b \over l_{\rm AdS}^2}
\right) + {\cal S}_0^{\rm AdS}\ .
\ee
Here $S_0^{\rm AdS}$ is a constant of the integration. 
If we assume that the entropy should vanish in the pure AdS, 
we have $S_0^{\rm AdS}=0$. 
On the other hand, when the spacetime is asymptotically dS, 
one has the following expressions for the thermodynamical energy
\be
\label{ABC11dS}
M_{\rm dS}=E_{\rm dS}
={3l_{\rm dS}^2 V_3 \over 16} \left({1 \over \kappa^2} 
+ {40a \over l_{\rm dS}^2} + {8b \over l_{\rm dS}^2} 
 + {4c \over l_{\rm dS}^2}\right)\left(4  
 - {16\mu_{\rm dS} \over l_{\rm dS}^2}\right)
\ee
and the entropy for $c=0$ case, 
\be
\label{lll6}
{\cal S}_{\rm dS}= {V_{3}\pi r_{\rm dS-csm}^3 \over 2}
\left( {8 \over \kappa^2} + {320 a + 64b \over l_{\rm dS}^2}
\right)+ {\cal S}_0^{\rm dS} \ .
\ee
Here $l_{\rm dS}^2=-l_{\rm AdS}^2$. Since the horizon radius 
$r_{\rm csm} $ corresponds to the cosmological horizon,  
 $r_{\rm dS-csm}^2=l_{\rm dS}^2$ for  pure de Sitter space. 
When the spacetime is asymptotically 
de Sitter, the integration constant ${\cal S}_0^{\rm dS}$ may 
be again determined by requiring that the entropy 
${\cal S}_{\rm dS}$ should vanish in the Nariai limit: 
\be
\label{dSAdS3}
S_0^{\rm dS}= -{V_{3}\pi l_{\rm dS}^3 \over 4\sqrt{2}}
\left( {8 \over \kappa^2} + {320 a + 64b \over l_{\rm dS}^2}
\right)
\ee
Substituting the solution (\ref{dSAdS2}), we find
\be
\label{dSAdS4}
{\cal S}_{\rm AdS,dS}= {V_{3}\pi r_{\rm AdS-bh,\,dS-csm}^3 
\over 2}
\left( - {16 \over \kappa^2} \mp \sqrt{{36 \over \kappa^4} 
+ \Lambda(80a + 16b)}\right) + {\cal S}_0^{\rm AdS,dS}\ .
\ee
Since the entropy should be positive, the lower sign in 
(\ref{dSAdS2}) or (\ref{dSAdS4}) should be chosen. Furthermore 
the condition that the entropy corresponding to the lower sign is 
really positive is
\be
\label{dsAdS5}
{20 \over \kappa^4\Lambda^2} + {80a + 16b \over \Lambda}
\geq 0\ .
\ee
In case that there are both of solutions which have positive 
entropy and negative one, the solution corresponding to 
negative entropy is unstable. The next interesting quantity is the free
energy $F$. By using the 
thermodynamical relation $F = E - T{\cal S}$, one obtains the 
following expression for the asymptotically AdS spacetime case:
\bea
\label{freeA}
F_{\rm AdS}&=& -V_{3} \left( {1 \over \kappa^2} 
 - {40 a +8b \over l_{\rm AdS}^2} \right) 
 \left( {r_{\rm AdS-bh}^4 \over l_{\rm AdS}^{2}}
 - {k \over 2}r_{\rm AdS-bh}^2 - {3l_{\rm AdS}^2 k^2 
 \over 16}\right) \nn
&& - \left( {r_{\rm AdS-bh} \over \pi l_{\rm AdS}^{2}}
 + {k \over 4\pi r_{\rm AdS-bh}}\right) {\cal S}_0^{\rm AdS} 
\eea
from (\ref{E6}), (\ref{E9}), (\ref{ABC11}) and (\ref{lll5}). 
On the other hand, for the asymptotically dS spacetime case, 
 using (\ref{E8}), (\ref{E10}), (\ref{ABC11dS}) and 
(\ref{lll6}) one gets 
\bea
\label{freedS}
F_{\rm dS}&=& -V_{3} \left( {1 \over \kappa^2} 
+ {40 a +8b \over l_{\rm dS}^2} \right) 
 \left( {r_{\rm dS-csm}^4 \over l_{\rm dS}^{2}}
+ r_{\rm dS-csm}^2 - {3l_{\rm dS}^2 \over 4} \right.\nn
&& \left. -{\sqrt{2}l_{\rm dS} \over r_{\rm dS-csm}} 
 \left( r_{\rm dS-csm}^2  - {l_{\rm dS}^{2} \over 2}
\right)\right)\ .
\eea
Then when the entropy vanishes, that is, ${1 \over \kappa^2} 
 - {40 a +8b \over l_{\rm AdS}^2}=0$ (for $S_0^{\rm AdS}=0$ 
 case) or ${1 \over \kappa^2} + {40 a +8b \over l_{\rm dS}^2}$, 
the free energy also vanishes. Then the negative entropy 
solution  corresponds not to the minimum but to the maximum 
of the free energy. In the path integral formulation, the free 
energy corresponds to the product of the temperature and 
the action $S$ with the minus sign: $F=-TS$. Therefore the 
minus entropy  corresponds to the minimum of the action. 
Therefore, this consideration indicates that the solution with negative
entropy 
is instable.
 
Let us evaluate the action by substituting the classical 
solution  (\ref{cvA}) into the expression (\ref{vi}) with 
$d+1=5$. We first consider about the de Sitter case before 
AdS case. Then  
\bea
\label{Action1}
S_{\rm dS}&=&V_3 \int_0^\beta dt 
\int_{r_{\rm dS-bh}}^{r_{\rm dS-csm}} dr r^3 \left(
{400 a + 80b \over l_{\rm dS}^4} + {20 \over 
\kappa^2 l_{\rm dS}^2} - \Lambda\right) \nn
&=& 2\beta V_3 \left(r_{\rm dS-csm}^4 - r_{\rm dS-bh}^4\right)
\left({40 a + 8b \over l_{\rm dS}^4} + {1 \over 
\kappa^2 l_{\rm dS}^2} \right) \nn
&=& 2\beta V_3 \left(2r_{\rm dS-csm}^2 - l_{\rm dS}^2\right)
\left( {1 \over 
\kappa^2} + {40 a + 8b \over l_{\rm dS}^2}\right) \ .
\eea
Here $\beta$ is the period of the time coordinate. In the second 
line, Eq.(\ref{llA}) with $d=4$ and $l_{\rm AdS}^2=-l_{\rm dS}^2$ 
is used. In the last line, Eqs.(\ref{E7}) and (\ref{E8}) are 
used. The action (\ref{Action1}) surely vanishes when the entropy 
vanishes $\left(\left({1 \over 
\kappa^2} + {40 a + 8b \over l_{\rm dS}^2} \right) =0\right)$. 
Since $2r_{\rm dS-csm}^2 - l_{\rm dS}^2>0$, the 
action is positive (negative) when the entropy is 
positive (negative) (the 
sign in front of the action has been chosen so that the 
partition function $Z$ is given by $Z\sim \e^S$). 
Therefore the solution with negative entropy should be really 
instable. 

In case that the spacetime is asymptotically anti-de Sitter, the 
action itself diverges due to the infinite volume of the AdS. 
In order to regularize the action, we cut off the spacetime at 
$r=r_{\rm max}$ and put the surface counterterms there\cite{NOOr2a}:
\bea
\label{Iiv}
S_b&=&S_b^{(1)} + S_b^{(2)} \nn
S_b^{(1)} &=& \int d^4 x \sqrt{- g}\left[
4 a \nabla_\mu n^\mu A + 2 b\left(n_\mu n_\nu \nabla_\sigma n^\sigma
 + \nabla_\mu n_\nu \right) B^{\mu\nu} \right. \nn
&& \left. + 8 c n_\mu n_\nu \nabla_\tau n_\sigma C^{\mu\tau\nu\sigma} 
+ {2 \over \kappa^2}\nabla_\mu n^\mu \right] \nn
S_b^{(2)} &=& - \int d^4 x \sqrt{- g_{(4)}} \left(\eta_1 + \eta_2 
R_{(4)}\right)\ .
\eea
Here $g_{(4)}$ is the metric induced on the boundary and 
$R_{(4)}$ is the scalar curvature given by $g_{(4)}$. 
$n^\mu$ is the normal outward unit vector to the boundary surface. 
$\eta_1$ and $\eta_2$ are constants given by
\be
\label{ABC8}
\eta_1={6 \over l_{\rm AdS}}\left({1 \over \kappa^2} 
 - {40a + 8b \over l_{\rm AdS} ^2} \right)\ ,\quad 
\eta_2=kl_{\rm AdS} \left({1 \over \kappa^2} 
 - {40a+8b \over l_{\rm AdS}^2} \right)\ .
\ee
Then the actions have the following forms:
\bea
\label{Action2}
S_{\rm AdS}&=& -2\beta V_3 \left(r_{\rm max}^4
 - r_{\rm AdS-bh}^4\right)
\left( {1 \over \kappa^2 l_{\rm AdS}^2} 
 - {40 a + 8b \over l_{\rm AdS}^4}\right) \ ,\\
\label{Action3}
S_b&=&\beta V_3 \e^{\nu(r_{\rm max})} 
\left( {1 \over \kappa^2 l_{\rm AdS}^2} 
 - {40 a + 8b \over l_{\rm AdS}^4}\right)r_{\rm max}^3 \nn
&& \times\left( - 2\nabla_\mu n^\mu + {6 \over l_{\rm AdS}} 
+ kl_{\rm AdS} R_{(4)}
\right) \ .
\eea
Then total action including the surface counterterms vanishes 
again when the entropy vanishes $\left(\left({1 \over 
\kappa^2} - {40 a + 8b \over l_{\rm AdS}^2} \right) =0\right)$.

\unitlength=0.45mm

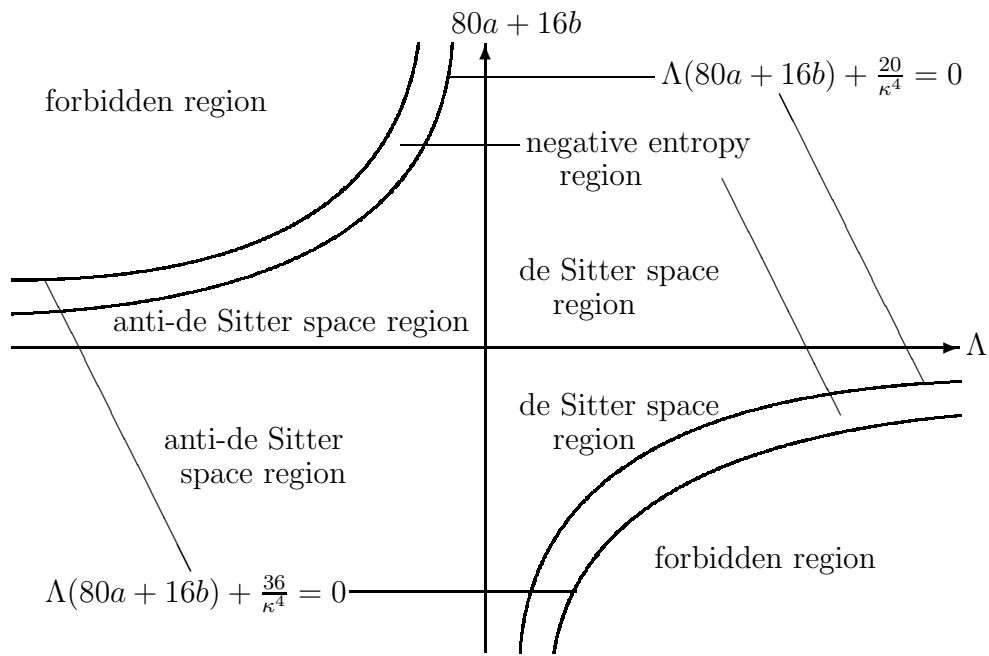
\begin{figure}
\begin{picture}(300,200)
\thicklines
\put(140,193){$80a+16b$}
\put(292,98){$\Lambda$}
\put(10,100){\vector(1,0){280}}
\put(150,10){\vector(0,1){180}}
\qbezier[300](10,110)(135,115)(140,190)
\qbezier[300](290,90)(165,85)(160,10)
\qbezier[300](10,120)(120,120)(130,190)
\qbezier[300](290,80)(180,70)(170,10)
\put(202,178){$\Lambda(80a + 16b) + {20 \over \kappa^4}=0$}
\put(20,25){$\Lambda(80a + 16b) + {36 \over \kappa^4}=0$}
\put(160,120){de Sitter space}
\put(170,110){region}
\put(160,80){de Sitter space}
\put(170,70){region}
\put(40,105){anti-de Sitter space region}
\put(55,70){anti-de Sitter}
\put(60,60){space region}
\put(162,158){negative entropy}
\put(172,148){region}
\put(20,170){forbidden region}
\put(200,35){forbidden region}
\thinlines
\put(139,180){\line(1,0){61}}
\put(280,89){\line(-1,2){43}}
\put(20,120){\line(1,-2){43}}
\put(177,28){\line(-1,0){67}}
\put(160,160){\line(-1,0){35}}
\put(220,150){\line(1,-2){35}}
\end{picture}
\caption{The relation between the parameters and the spacetime structure}
\label{Fig1}
\end{figure}

The obtained results are summarized in Figure \ref{Fig1}:
The parameters region outside the curve $\Lambda(80a + 16b) 
+ {36 \over \kappa^4}=0$ is presumably forbidden since 
$l_{\rm AdS}^2$ becomes 
complex number there. (Nevertheless, in principle there is possibility 
of realization of real well-defined four-dimensional brane even in the
situation
when 5d bulk space has complex curvature). In the region between two curves, 
$\Lambda(80a + 16b) + {20 \over \kappa^4}=0$ and $\Lambda(80a + 16b) 
+ {36 \over \kappa^4}=0$, the entropy is always negative. As it follows from
the action principle  these 
regions are not realistic as corresponding solutions are not stable.
Asymptotically anti-de Sitter 
(de Sitter) spacetime with $l_{\rm AdS}^2>0$ ($l_{\rm AdS}^2<0$) 
can appear as 
a solution of Eq.(\ref{llA4}) even if $\Lambda>0$ ($\Lambda<0$). 
These solutions have 
always negative entropy and again are not stable as it was shown above. As
a result, 
as in the Einstein theory, the anti-de Sitter (de Sitter) 
spacetime corresponds to the negative (positive) $\Lambda$. 

\section{The brane motion and dual CFTs}

Having in mind AdS/CFT and dS/CFT correspondence, 
one can investigate the structure in Fig.\ref{Fig1} from 
the dual CFT side. For this purpose, we consider the motion of 
the brane in the Schwarzschild-AdS or dS black hole 
spacetime. We call that the brane is time (space)-like if the 
vector normal to the brane is space (time)-like. 
We now assume the brane exists at $r=\tilde a(\tau)$, 
where $\tau$ is the proper time on the brane. Then 
the Hubble ``constant''  $H$ is defined by
\be
\label{FRW1}
H={1 \over \tilde a}{d\tilde a \over d\tau}\ ,
\ee
the motion of the time-like brane obeys the following equation
\be
\label{FRW2}
H^2={1 \over l_{\rm dS,AdS}^2} 
 - {\e^{2\rho\left(r=\tilde a(\tau)\right)} \over \tilde a^2}\ ,
\ee
and that of the space-like brane obeys \cite{ogushi}
\be
\label{FRW3}
H^2={1 \over l_{\rm dS,AdS}^2}
+ {\e^{2\rho\left(r=\tilde a(\tau)\right)} \over \tilde a^2}\ .
\ee
Then in case of the SAdS background (\ref{E2}), one obtains
\bea
\label{FRW4}
&& H^2=- {k \over 2\tilde a^2} 
+ {\mu_{\rm AdS} \over \tilde a^4}\ \mbox{(for the time-like 
brane)} \nn
&& \mbox{or}\ H^2={2 \over l_{\rm AdS}^2} + {k \over 2\tilde a^2} 
 - {\mu_{\rm AdS} \over \tilde a^4}
\ \mbox{(for the space-like brane)} \ ,
\eea
and in case of the SdS background (\ref{E4})
\bea
\label{FRW5}
&& H^2={2 \over l_{\rm dS}^2}- {1 \over \tilde a^2} 
+ {\mu_{\rm dS} \over \tilde a^4}\ 
\mbox{(for the time-like brane)} \nn
&& \mbox{or}\ 
H^2=  {1 \over \tilde a^2} - {\mu_{\rm dS} \over \tilde a^4}
\ \mbox{(for the space-like brane)} \ .
\eea
The above equations can be rewritten in the form of the 
FRW equation:
\bea
\label{FRW6}
&& H^2 = - {k \over 2\tilde a^2} + {\kappa_4^2 \over 6}
{\tilde E_{\rm AdS} \over V}\ 
\mbox{(for the time-like brane)} \nn
&& \mbox{or}\ 
H^2 = {2 \over l_{\rm AdS}^2} + {k \over 2\tilde a^2} 
 - {\kappa_4^2 \over 6} {\tilde E_{\rm AdS} \over V}\ 
\mbox{(for the space-like brane)} \ ,\nn
&& {\tilde E_{\rm AdS}}={6 \mu_{\rm AdS} V_3 \over 
\kappa_4^2 \tilde a}\ ,\quad V=\tilde a^3 V_3
\eea 
for the brane in SAdS and 
\bea
\label{FRW7}
&& H^2 = {2 \over l_{\rm dS}^2} 
 - {1 \over \tilde a^2} + {\kappa_4^2 \over 6}
{\tilde E_{\rm dS} \over V}\ 
\mbox{(for the time-like brane)} \nn
&& \mbox{or}\ 
H^2 = {1 \over \tilde a^2} - {\kappa_4^2 \over 6}
{\tilde E_{\rm dS} \over V}\ \mbox{(for the space-like brane)} \ ,\nn
&& {\tilde E_{\rm dS}}={6 \mu_{\rm dS} V_3 \over 
\kappa_4^2 \tilde a}\ ,
\eea
for the brane in SdS. 
Here $V_3$ is the volume of the three--dimensional sphere 
with  unit radius  and $\kappa_4$ is the four--dimensional 
gravitational coupling. 
Usual choice for effective gravitational constant in higher derivative
gravity is
\cite{NOOr2a,cvetic,neupane}
\bea
\label{FRW8}
&& {1 \over \kappa_4^2} = {1 \over \tilde \kappa_{\rm AdS}^2}
\equiv {l_{\rm AdS} \over 2}\left({1 \over \kappa^2}
 - {40 a \over l_{\rm AdS}^2} -{8 b \over l_{\rm AdS}^2} \right) \nn
&&  \mbox{or}\ 
{1 \over \kappa_4^2} = {1 \over \tilde \kappa_{\rm dS}^2}
\equiv {l_{\rm dS} \over 2}\left({1 \over \kappa^2}
 + {40 a \over l_{\rm dS}^2} + {8 b \over l_{\rm dS}^2} \right)\ ,
\eea 
but Eq.(\ref{FRW8}) can be a convention at this stage. 
The motion of the brane 
itself does not depend on the explicit definition of $\kappa_4^2$. Then 
even if the black hole entropy vanishes at ${1 \over \kappa^2}
 - {40 a \over l_{\rm AdS}^2} -{8 b \over l_{\rm AdS}^2} =0$ 
 or ${1 \over \kappa^2} + {40 a \over l_{\rm dS}^2} 
 + {8 b \over l_{\rm dS}^2} =0$, the motion of the brane is 
not singular. 

By differentiating Eqs.(\ref{FRW6}) and (\ref{FRW7}), we 
obtain the second FRW equation:
\be
\label{FRW14}
\dot H = \mp {\kappa_4^2 \over 4} \left({\tilde E_{\rm dS,AdS} 
\over V} + p_{\rm dS,AdS}\right) \pm {k \over 2\tilde a^2}\ ,\quad
p_{\rm dS,AdS}={2\mu_{\rm dS,AdS} \over \tilde a^4 
\kappa_4^2}\ .
\ee 
Here $k=2$ for the bulk dS and the upper sign corresponds to 
the time-like brane and the lower sign to the space-like brane. 
Since $3p_{\rm dS,AdS}=\tilde E_{\rm dS,AdS}/V$, the trace of the 
energy-stress tensor arising from the matter on the brane 
vanishes, i.e.,  ${T^{{\rm matter}\ \mu}}_\mu=0$. 
Thus, the matter on the brane can be regarded as radiation or, 
equivalently, as massless fields. In other 
words, dual field theory on the brane should be a conformal theory. 

We now compare the brane energy ${\tilde E_{\rm AdS}}$ 
or ${\tilde E_{\rm dS}}$  
(\ref{FRW6}) or (\ref{FRW7}) with the 5d black hole mass  
(\ref{ABC11}) or (\ref{ABC11dS}). In case of the bulk AdS, 
if one subtracts the bulk Casimir energy 
\be
\label{EE1}
E_{\rm AdS\, C} 
= {3l_{\rm AdS}^2 V_3 \over 16} \left({1 \over \kappa^2} 
 - {40a \over l_{\rm AdS}^2}  - {8b \over l_{\rm AdS}^2}\right)k^2 
={3l_{\rm AdS} V_3 \over 8} 
{k^2 \over \tilde \kappa_{\rm AdS}^2}\ ,
\ee
from the black hole mass, one gets 
\be
\label{EE2}
E_{\rm AdS} - E_{\rm AdS\, C} 
= {l_{\rm AdS} \over \tilde a}
{\kappa_4^2 \over \tilde \kappa_{\rm AdS}^2}
\tilde E_{\rm AdS}\ .
\ee
Since $E_{\rm AdS\, C}$ is independent of the parameters 
characterizing the black hole, for example, of the radius of 
the black hole horizon, $E_{\rm AdS\, C}$ would be proper characteristic 
of the 
background where the black hole lies. Therefore one can 
regard that $E_{\rm AdS\, C}$ should be a Casimir energy 
of AdS. 
Since the ratio of the time on the brane with the bulk 
time is given by ${l_{\rm AdS} \over \tilde a}$ when the radius 
of the brane is large, it would be consistent the following choice 
$\kappa_4^2 = \tilde \kappa_{\rm AdS}^2$. Then when 
the black hole entropy vanishes at ${1 \over \kappa^2}
 - {40 a \over l_{\rm AdS}^2} -{8 b \over l_{\rm AdS}^2} =0$, 
the energy of the CFT on the brane also vanishes. 
And if the entropy is negative, the energy is also negative. 
In the following, we choose $\kappa_4^2 
= \tilde \kappa_{\rm AdS}^2$ for the case that the spacetime 
is asymptotically AdS.  

On the other hand, in case of the bulk dS, even 
if we subtract the bulk Casimir energy 
\be
\label{EE3}
E_{\rm dS\, C} 
= {3l_{\rm dS}^2 V_3 \over 4} \left({1 \over \kappa^2} 
+ {40a \over l_{\rm dS}^2} + {8b \over l_{\rm dS}^2} \right) 
={3l_{\rm dS} V_3 \over 2} {1 \over \tilde \kappa_{\rm AdS}^2}\ ,
\ee
from the black hole mass, we find 
\be
\label{EE4}
E_{\rm dS} - E_{\rm dS\, C} = -{l_{\rm dS} \over \tilde a}
{\kappa_4^2 \over \tilde \kappa_{\rm dS}^2}
\tilde E_{\rm dS}\ .
\ee
Therefore there is a difference in sign. 
Then we should not identify the energy of the CFT on the brane 
with $\tilde E_{\rm dS}$ but with 
\be
\label{hatE}
\hat E_{\rm dS} \equiv - \tilde E_{\rm dS}\ . 
\ee
For this choice, we may choose 
$\kappa_4^2 = \tilde \kappa_{\rm dS}^2$. Then when 
the black hole entropy vanishes at ${1 \over \kappa^2}
+ {40 a \over l_{\rm dS}^2} + {8 b \over l_{\rm dS}^2} =0$, 
the energy of dual CFT on the brane also vanishes. 
When the entropy is positive (negative), the energy is 
also positive (negative). 
In the following, we choose $\kappa_4^2 
= \tilde \kappa_{\rm dS}^2$ for the bulk dS.

%%%%%%%%%%%%%%%%%%%%%%%%
In case of the bulk AdS, one assumes that the total 
entropy ${\cal S}_{\rm AdS}$ of the CFT on the brane is given 
by Eq. (\ref{lll5}) with ${\cal S}_0^{\rm AdS}=0$. 
Then if this entropy is constant during the 
cosmological evolution, the entropy density $s_{\rm AdS}$ is 
given by 
\be
\label{abe20}
s_{\rm AdS}={{\cal S}_{\rm AdS} \over \tilde a^3 V_3}
={8\pi r_{\rm AdS-bh}^3 \over l_{\rm AdS} 
\tilde \kappa_{\rm AdS}^2 \tilde a^3}\ .
\ee
If we further assume that the brane temperature $T$  
differs from the Hawking temperature $T_{\rm AdS-bh}$  
(\ref{E9}) (the expression does not change  for the 
$R^2$-gravity with $c=0$) by the factor $l_{\rm AdS}/{\tilde a}$, 
one finds 
\be
\label{abe22}
T={l_{\rm AdS} \over \tilde a}T_{\rm AdS-bh}
={r_{\rm AdS-bh} \over \pi \tilde a l_{\rm AdS}} 
+ {kl_{\rm AdS} \over 4\pi \tilde a r_{\rm AdS-bh}} 
\ee
and, when $\tilde a=r_{\rm AdS-bh}$, this implies that  
\be
\label{abe23}
T={1 \over \pi l_{\rm AdS}} 
+ {k \over 4\pi r_{\rm AdS-bh}^2}\ .
\ee
If the energy and entropy are purely extensive, the 
quantity $\tilde E_{\rm AdS} + p_{\rm AdS}V 
 - T{\cal S}_{\rm AdS}$ vanishes. In general, 
this condition does not hold and one can define the CFT 
Casimir energy $\tilde E_{\rm AdS\, C}$ (we should distinguish 
it with the bulk Casimir energy  (\ref{EE1})) as \cite{verlinde} 
\be
\label{abEC1}
\tilde E_{\rm AdS\, C}=3\left(\tilde E_{\rm AdS} + p_{\rm AdS} V 
 - T{\cal S}_{\rm AdS}\right)\ .
\ee
Then 
\be
\label{abEC2}
\tilde E_{\rm AdS\, C}={6k r_{\rm AdS-bh}^2 V \over \kappa_4^2 
\tilde a^4} 
= {6k r_{\rm AdS-bh}^2 V_3 \over \kappa_4^2 \tilde a} \ .
\ee
If the black hole entropy vanishes at 
${2 \over l_{\rm AdS}\kappa_4^2} 
= {2 \over l_{\rm AdS} \tilde \kappa_{\rm AdS}^2}
= {1 \over \kappa^2}
 - {40 a \over l_{\rm AdS}^2} -{8 b \over l_{\rm AdS}^2} =0$, 
the Casimir energy of the CFT on the brane also vanishes. 
Furthermore, when the black hole entropy is negative, 
the Casimir energy is negative (positive) if $k>0$ ($k<0$). 
When $k\neq 0$, we also find that
\be
\label{abSS}
{\cal S}_{\rm AdS}^2=\left({4\pi \tilde a \over 3\sqrt{|2k|}}\right)^2
\tilde E_{\rm AdS\, C}\left(2\tilde E_{\rm AdS} 
 - \tilde E_{\rm AdS\, C}\right)\ .
\ee
This is the form of Cardy-Verlinde formula in AdS higher 
derivative gravity
under consideration.

In case that the bulk is dS, the above arguments should be 
modified. First one needs to assume that the total entropy 
of the CFT on the brane is not given by Eq. (\ref{lll6}) but 
by the black hole (not cosmological) horizon area
\be
\label{abe20dS}
{\cal S}_{\rm dS}^p 
={8\pi r_{\rm dS-bh}^3 V_3 \over l_{\rm dS}
\tilde \kappa_{\rm dS}^2}\ .
\ee
We also assume that the temperature $T$ on the brane 
is given, in terms of the black hole (not cosmological) 
horizon radius, by
\be
\label{abe22dS}
T={l_{\rm dS} \over \tilde a}T_{\rm dS-bh}
=-{r_{\rm dS-bh} \over \pi \tilde a l_{\rm dS}} 
+ {l_{\rm dS} \over 2\pi \tilde a r_{\rm dS-bh}} \ .
\ee
Then the CFT Casimir energy $\tilde E_{\rm dS\, C}$ defined by 
\be
\label{abEC1dS}
\tilde E_{\rm dS\, C}=3\left(\tilde E_{\rm dS} + p_{\rm dS} V 
 - T{\cal S}_{\rm dS}^p\right)\ .
\ee
has the following form
\be
\label{abEC2dS}
\tilde E_{\rm dS\, C}={12 r_{\rm dS-bh}^2 V \over 
\kappa_4^2 \tilde a^4} \ .
\ee
If the black hole entropy vanishes at 
${2 \over l_{\rm dS}\kappa_4^2} 
= {2 \over l_{\rm dS} \tilde \kappa_{\rm dS}^2}
= {1 \over \kappa^2}
+ {40 a \over l_{\rm dS}^2} + {8 b \over l_{\rm dS}^2} =0$, 
the Casimir energy of the CFT on the brane also vanishes. 
One also arrives at Cardy-Verlinde formula in dS brane-world
\be
\label{abSSdS}
\left({\cal S}_{\rm dS}^p\right)^2
=\left({2\pi \tilde a \over 3}\right)^2
\tilde E_{\rm dS\, C}\left(2\tilde E_{\rm dS} - \tilde E_{\rm dS\,
C}\right)\ .
\ee
The above formula (\ref{abSSdS}) seems to be identical 
with the corresponding expression (\ref{abSS}) for AdS bulk 
since $|k|=2$. Nevertheless, the meaning of the entropy and the energy 
seems to be different. In case of SAdS, these quantities 
correspond to those of the whole spacetime of SAdS but 
in case of AdS they only correspond to the quantities 
proper for the black hole. 

If, instead of (\ref{abe20dS}) and (\ref{abe22dS}),
we define the entropy and the temperature in terms of the 
radius of the cosmological horizon:
\be
\label{abe20dSB}
{\cal S}_{\rm dS}^c 
={8\pi r_{\rm dS-csm}^3 V_3 \over l_{\rm dS}
\tilde \kappa_{\rm dS}^2}\ ,\quad 
T^c={l_{\rm dS} \over \tilde a}T_{\rm dS-csm}
={r_{\rm dS-csm} \over \pi \tilde a l_{\rm dS}} 
 - {l_{\rm dS} \over 2\pi \tilde a r_{\rm dS-csm}} \ ,
\ee
where ${\cal S}_{\rm dS}^c$ corresponds to Eq. (\ref{lll6}) with 
${\cal S}^{\rm dS}_0=0$, then by using $\hat E_{\rm dS}$ in 
(\ref{hatE}) instead of $\tilde E_{\rm dS}$, 
the Casimir energy has the following form:
\be
\label{abEC1dSB}
\tilde E_{\rm dS\, C}^c=3\left(\hat E_{\rm dS} + p_{\rm dS} V 
 - T^c{\cal S}_{\rm dS}^c\right)
= - {12 r_{\rm dS-bh}^2 V \over \kappa_4^2 \tilde a^4} \ .
\ee
Then, one gets 
\be
\label{abSSdSB}
\left({\cal S}_{\rm dS}^c\right)^2
=\left({2\pi \tilde a \over 3}\right)^2 \tilde E_{\rm dS\, C}^c
\left(2\hat E_{\rm dS} + \tilde E_{\rm dS\, C}^c\right)\ .
\ee
The sign in front of $\tilde E_{\rm dS\, C}^c$ in the 
last factor is different from (\ref{abSSdS}). 
However, the physical meaning of this relation is not clear.
The conjecture could be that there are two types of dual CFT 
for SdS BH.

Thus, in this section it is considered the brane motion on (A)dS bulk 
and it is found the relation between energy, Casimir energy and entropy of
dual CFT. 

\section{Discussion}
In summary, the investigation of thermodynamics of five-dimensional (A)dS
BHs in higher derivative gravity is presented. Entropy, free energy and mass
(thermodynamical energy) for (A)dS BHs is evaluated and their properties 
are discussed.
It is shown that there exists the parameters region where the effective 
gravitational constant is getting negative (zero) and, as a result, the
entropy becomes negative (zero) too. The arguments based on least action
principle are given that  negative entropy BHs are  instable and their
creation is highly suppressed. It is interesting to note that there exists 
RG interpretation for classical solutions of higher derivative gravity
\cite{fukuma}. It could be that negative entropy BHs are not stable also
from the point of view of holographic RG.

The motion of time (space)-like branes in SAdS or SdS BH background is
considered. The corresponding equations of motion are presented in
FRW-form. The brane motion is not singular even when effective
gravitational constant (BH entropy) is zero. It is demonstrated that dual
QFT for 
higher derivative gravity model under consideration is CFT.
Moreover, when BH entropy is zero the brane CFT energy is also zero.
Finally, Cardy-Verlinde formula relating energy, entropy and Casimir energy 
of dual CFT is written in case of SAdS as well as SdS BHs. 

Our study indicates that there are deep similarities between SAdS and SdS
BHs and corresponding dual CFTs living on the branes. In particulary, 
Cardy-Verlinde formula looks the same in both cases, the BH entropy 
may be
negative (zero) in both cases, etc. In this sense, our study provides
further support to dS/CFT correspondence. 
As (A)dS BH thermodynamics is well-described here even in higher
derivative gravity case, the natural next problem is the investigation of
the properties 
of dual CFT. This will be discussed elsewhere.

\section*{Acknowledgements} 

The authors would like to acknowledge helpful discussions with S.Ogushi.
The work by SN is supported in part by the Ministry of Education, 
Science, Sports and Culture of Japan under the grant No. 13135208.


\begin{thebibliography}{99}
\bibitem{AdS} J.M. Maldacena, {\sl Adv.Theor.Math.Phys.} {\bf 2} (1998) 231;
E. Witten, {\sl Adv.Theor.Math.Phys.} {\bf 2} (1998) 253;
S. Gubser, I. Klebanov and A. Polyakov, {\sl Phys.Lett.} 
{\bf B428} (1998) 105, 
O. Aharony, S. Gubser, J. Maldacena, H. Ooguri and Y. Oz,
{\it Phys. Rept. \/} {\bf 323} 183, (2000), hep-th/9905111.
\bibitem{strominger} A. Strominger, hep-th/0106113;
M. Spradlin, A. Strominger and A. Volovich, hep-th/0110007.
\bibitem{hull} C.M. Hull, {\sl JHEP} {\bf 9807} (1998) 021, 
hep-th/9806146.
\bibitem{ds2} E. Witten, hep-th/0106109.
\bibitem{ds3a}
P.O. Mazur and E. Mottola, hep-th/0106151;
M. Li, hep-th/0106184;
S. Nojiri and S.D. Odintsov,
{\sl Phys.Lett.} {\bf B519} (2001) 145, hep-th/0106191;
{\sl JHEP} {\bf 0112} (2001) 033, hep-th/0107134; 
{\sl Phys.Lett.} {\bf B523} (2001) 165, hep-th/0110064; 
E. Halyo, hep-th/0107169;
C.M. Hull, hep-th/0109213;
T. Shiromizu, D. Ida and T. Torii, hep-th/0109057;
S. Cacciatori and D. Klemm, hep-th/0110031;
B. McInnes, hep-th/0110062;
Y.Gao, hep-th/0107067;
I.Y. Park, C.N. Pope and A. Sadrzadeh, hep-th/0110238;
U. Danielsson, hep-th/0110265;
Y. Myung, hep-th/0201176, hep-th/0112140;
R.-G. Cai, Y. Myung and Y. Zhang, hep-th/0110234;
R.-G. Cai, hep-th/0111093;
A.C. Petkou and G. Siopsis, hep-th/0111085;
A. Padilla, hep-th/0111247;
A. Medved,  hep-th/0111238;
D. Youm, hep-th/0111276; B. Cunha, hep-th/0110169;
M.Li and F. Lin, hep-th/0111201;
M. Dehghani,  hep-th/0201128;
R. Bousso, A. Maloney and A. Strominger, hep-th/0112218;
M. Spradlin and A. Volovich, hep-th/0112223;
A. Ghezelbash, D. Ida, R. Mann and T. Shiromizu, hep-th/0201004;
E. Halyo, hep-th/0201174, hep-th/0203235.
\bibitem{ogushi} S. Ogushi, {\sl Mod.Phys.Lett.} {\bf A17} 
(2002) 51, hep-th/0111008. 
\bibitem{others} D. Kabat and G. Lifschytz, hep-th/0203083;
G. Siopsis, hep-th/0203208;
S.R. Das, hep-th/0202008;
S. Ness and G. Siopsis, hep-th/0202096;
S. Nojiri, S.D. Odintsov and S. Ogushi, hep-th/0202098;
A. Ghezelbash and R.B. Mann, hep-th/0203003;
A. Medved, hep-th/0202193;
D. Klemm and L. Vanzo, hep-th/0203268;
I. Brevik, K. Ghoroku, S.D. Odintsov and M. Yahiro, hep-th/0204066;
S. Nojiri, S.D. Odintsov and A. Sugamoto, hep-th/0204065;
H. Lee and Y. Myung, hep-th/0204083.
\bibitem{cvetic} M. Cveti\v c, S. Nojiri, and S. D. Odintsov,
to appear in {\sl Nucl.Phys.} {\bf B}, hep-th/0112045;
 S. Nojiri and S.D. Odintsov, 
gr-qc/0112066, to appear in the Proceedings of 
the ``1st Mexican Meeting on Mathematical and 
Experimental Physics", held at El Colegio Nacional, 
10-14 September, 2001;
J.E. Lidsey, S. Nojiri and S.D. Odintsov, hep-th/0202198.
\bibitem{NOOr2a} S. Nojiri, S.D. Odintsov and S. Ogushi, 
hep-th/0105117, {\sl Int.J.Mod. Phys.} {\bf A16} (2001) 5085;
hep-th/0108172, {\sl Phys.Rev.} {\bf D65} (2001) 023521.
\bibitem{BOS} I.L. Buchbinder, S.D. Odintsov and I.L. Shapiro,
``Effective Action in Quantum Gravity'', IOP Publishing,
Bristol and Philadelphia, 1992.
\bibitem{BK} V. Balasubramanian and P. Kraus, 
{\sl Commun.Math.Phys.} {\bf 208} (2000) 413. 
\bibitem{BBM} V. Balasubramanian, J. de Boer and D. Minic, 
hep-th/0110108.
\bibitem{neupane} Y.M. Cho and I.P. Neupane, hep-th/0202140.
\bibitem{verlinde} E. Verlinde, hep-th/0008140.
\bibitem{fukuma} M. Fukuma and S. Matsuura, hep-th/0112037.



\end{thebibliography}
\end{document}